\documentclass[a4paper, amsfonts, amssymb, amsmath, reprint, showkeys, nofootinbib, twoside,notitlepage,onecolumn]{revtex4-1}

\def\prd{{Phys.~Rev.~D}}        
\usepackage{amsmath,amstext}
\usepackage[T1]{fontenc}
\usepackage{amssymb}
\usepackage{graphicx}
\usepackage{ae,aecompl}

\DeclareFontFamily{OT1}{pzc}{}
\DeclareFontShape{OT1}{pzc}{m}{it}{<-> s * [1.10] pzcmi7t}{}
\DeclareMathAlphabet{\mathpzc}{OT1}{pzc}{m}{it}

\usepackage{hyperref}
\usepackage{amsmath}
\usepackage{amssymb}
\usepackage{mathtools}
\usepackage{bm}
\usepackage{cleveref}
\usepackage{tensor}
\usepackage{braket}
\usepackage{enumitem}
\usepackage{mhchem}
\usepackage{amsthm}
\usepackage{nccmath}
\usepackage{mathrsfs}
\usepackage{color}

\newcommand{\Tt}{{\Tilde{t}}}
\newcommand{\Tx}{{\Tilde{x}}}
\newcommand{\Tk}{{\Tilde{k}}}
\newcommand{\Tom}{{\Tilde{\omega}}}
\newcommand{\feq}{{f_{\mathrm{eq}}}}
\newcommand{\V}{{\mathscr{V}}}

\def\be{\begin{equation}}
\def\ee{\end{equation}}
\def\beq{\begin{eqnarray}}
\def\eeq{\end{eqnarray}}

\theoremstyle{definition}

\theoremstyle{theorem}

\theoremstyle{corollary}

\begin{document}
\title{Lorentz-boosted diffusion: initial value formulation and exact solutions}
\author{L.~Gavassino}
\affiliation{Department of Applied Mathematics and Theoretical Physics, University of Cambridge, Wilberforce Road, Cambridge CB3 0WA, United Kingdom}

\begin{abstract}
It is well known that the diffusion equation, when treated as a stand-alone partial differential equation, exhibits exponential instabilities in boosted frames, which render the corresponding initial-value problem ill-posed. Recently, however, it was shown that Fick-type diffusion arises as the exact hydrodynamic sector of relativistic Fokker-Planck kinetic theory. In this work, we exploit this kinetic embedding to formulate a modified initial-value problem for one-dimensional Lorentz-boosted diffusion. We show that the resulting dynamics are well posed both forward and backward in time, provided the boosted density profiles admit a kinetic-theory realization. Such profiles form a space of band-limited functions, within which the evolution can be expressed as a discrete superposition of spatially sampled initial data, weighted by a Shannon-Whittaker-type Green function defined on the full Minkowski plane. The Green function is obtained in closed analytic form.
\end{abstract}
 
\maketitle

\section{Introduction}

Let $n(t,x)$ denote the density of an ensemble of relativistic particles constrained to move in one spatial dimension. Such particles are assumed to propagate through a medium at rest, which induces stochastic scattering and gives rise to diffusive transport. In the long-wavelength limit, the density field is expected to obey Fick’s law \cite[\S 9.3]{peliti_book},
\begin{equation}\label{diffusion}
\partial_t n = \partial_x^2 n \, ,
\end{equation}
written here in natural units. A fundamental question in relativistic hydrodynamics \cite{Hiscock_Insatibility_first_order,Kost2000,Basar:2024qxd} is how such a diffusive description should be formulated in a reference frame $(\Tt,\Tx)$ in which the medium moves with constant velocity $v$, i.e.
\vspace{-0.2cm}
\begin{equation}
\begin{cases}
\Tt = \gamma (t + v x),\\
\Tx = \gamma (x + v t),\\
\end{cases} 
\end{equation}
with $\gamma = (1 - v^2)^{-1/2}$. Below, we briefly summarize the most natural (and common) strategies, and the difficulties associated with each of them.

A first possibility is to solve \eqref{diffusion} for prescribed initial data at $t=0$, and then to apply a Lorentz boost to the resulting solution. This procedure has the limitation that, for generic initial data, solutions of \eqref{diffusion} cannot be extended to negative times \cite[\S 2.1]{Isakov2006InverseProblems}. As a consequence, the boosted density profiles are typically defined only in the region $\Tt > v \Tx$ (corresponding to $t>0$). In particular, there exists no time $\Tt$ at which the density profile is defined for all spatial points $\Tx$. One may attempt to circumvent this issue by restricting the space of admissible initial data so that the corresponding solutions $n(t,x)$ exist for all $t\in\mathbb{R}$, but such a restriction appears physically unmotivated. For instance, it would exclude compactly supported initial data, which are otherwise admissible.

A second possibility is to boost equation \eqref{diffusion} itself,
\begin{equation}\label{diffusionBoosted}
(\partial_\Tt + v \partial_\Tx) n
=
\gamma (\partial_\Tx + v \partial_\Tt)^2 n \, ,
\end{equation}
and to solve it for initial data at $\Tt=0$. This approach encounters a more severe obstruction: the resulting initial-value problem is ill-posed in the sense of Hadamard, in that solutions fail to exist for generic Sobolev initial data \cite{Kost2000}. The origin of this pathology lies in the presence of exponential instabilities (e.g., $n= e^{\Tt/(\gamma v^2)}$ is a solution), whose growth rate diverges in the short-wavelength limit. For this reason, \eqref{diffusionBoosted} is commonly regarded as devoid of predictive content.

A third line of attack, widely explored in the literature, consists in modifying Fick’s law itself, so as to restore compatibility with relativistic requirements. In particular, guaranteeing stability in all reference frames necessitates rendering \eqref{diffusion} causal \cite{GavassinoSuperlum2021}, which can be achieved by the inclusion of a second-order time derivative \cite{cattaneo1958,Israel_Stewart_1979,GavassinoFronntiers2021,GavassinoAntonelli:2025umq}, as in $\partial_t n \,{=}\, (\partial_x^2 {-} \partial_t^2) n$. The resulting model, commonly referred to as Cattaneo theory \cite[\S 6.5.1]{rezzolla_book}, can indeed be shown (see Appendix \ref{aaa}) to reproduce the \emph{exact} density dynamics in a specific microscopic setting, namely one in which the particles are massless, and their velocity ($\V=\pm 1$) undergoes random sign flips with a momentum-independent rate \cite{Basar:2024qxd}. The drawback of this construction is that Cattaneo's theory lacks the universal character of Fick’s law \cite{Geroch1995,LindblomRelaxation1996,GerochCriticism:2001xs}.
Motivated by this limitation, a different strategy has recently been proposed \cite{Basar:2024qxd}, in which \eqref{diffusion} is left unchanged, but the Lorentz transformation is implemented only approximately. This procedure preserves the long-wavelength, low-frequency behavior of the theory, while reducing the temporal order of \eqref{diffusionBoosted} so as to guarantee well-posedness, leading to $(\partial_\Tt + v \partial_\Tx) n = \gamma^{-3} \partial_\Tx^2 n$. The price paid in this approach is the explicit loss of Lorentz covariance: different observers no longer construct solutions that are related to one another by Lorentz transformations \cite{GavassinoParabolic2025hwz}.

In this article, we explore yet another approach. We will start from the recent realization \cite{GavassinoDiffusionCompatible2026tvy,GavassinoFokkerPlanck2026zsz} that, when the microscopic dynamics are governed by Fokker-Planck momentum diffusion, the density $n$ obeys Fick’s law \eqref{diffusion} \emph{exactly} within the hydrodynamic sector, understood as the branch of modes continuously connected to the origin in frequency space. As a consequence, boosting relativistic Fokker-Planck kinetic theory (which is itself covariantly stable \cite{GavassinoDistrubingMoving:2026klp}) results in a hydrodynamic sector governed by \eqref{diffusionBoosted}. Importantly, however, not every solution of \eqref{diffusionBoosted} admits a realization within the kinetic theory, as exponentially unstable solutions such as $n = e^{\Tt/(\gamma v^2)}$ are automatically excluded. This observation provides a physically motivated criterion for selecting admissible initial data for boosted diffusion.
We exploit this criterion to derive an initial-value formulation of \eqref{diffusionBoosted} restricted to a function space that is substantially smaller than the Sobolev spaces conventionally employed in the theory of partial differential equations \cite{evansPDEbook}. We then find that such initial-value problem is well-posed both forward and backward in time, and that its solutions admit an explicit representation in terms of a discrete Green-function expansion.

Throughout the article, we work in natural units, $c=\hbar=k_B=\text{Diffusivity}=1$.

\vspace{-0.4cm}
\section{Boosted diffusion from Fokker-Planck kinetic theory}
\vspace{-0.3cm}

In this section, we show how the embedding of the diffusion equation into relativistic Fokker-Planck kinetic theory, introduced in \cite{GavassinoDiffusionCompatible2026tvy,GavassinoFokkerPlanck2026zsz}, can be exploited to identify a physically motivated reduced space of functions within which diffusion admits a well-posed formulation. Physically, Fokker-Planck describes particles propagating through a medium and subject to infinitely frequent but infinitely soft Markovian scattering, so that the particle momenta undergo Brownian motion. This is a fundamentally different limit from the standard relaxation-time approximation (RTA), where particles propagate ballistically over a finite timescale before undergoing fully randomizing scattering.

\vspace{-0.3cm}
\subsection{Diffusion as a sector of relativistic Fokker-Planck theory: brief recap}
\vspace{-0.3cm}

Let $p$ denote the momentum of an individual particle, $\varepsilon(p)$ its energy, and $\V = d\varepsilon/dp$ its velocity, all defined in the global rest frame of the external medium. The state of an ensemble of particles is described by the kinetic distribution function $f(t,x,p)$ \cite[\S 3.1]{huang_book}. In our units, the relativistic Vlasov-Fokker-Planck equation takes the form \cite{Debbasch,DunkelHanggi}
\begin{equation}\label{fokkerplanck}
(\partial_t + \V \partial_x) f
=
\beta^{-2}\partial_p\!\left[e^{-\beta\varepsilon}\,\partial_p\!\left(e^{\beta\varepsilon} f\right)\right],
\end{equation}
where $\beta>0$ denotes the inverse temperature of the external medium (which is assumed constant). Within this framework, the hydrodynamic sector is spanned by modes of the form
\begin{equation}\label{ansatzsolution}
f = e^{-\beta\varepsilon + i k (x - \beta p) - i \omega t} \, ,
\end{equation}
with $k,\omega \in \mathbb{C}$. In fact, substituting the ansatz \eqref{ansatzsolution} into \eqref{fokkerplanck}, one finds that the kinetic equation is satisfied if and only if $\omega = - i k^{2}$, which coincides with the dispersion relation of the diffusion equation \eqref{diffusion}. It follows immediately that the associated density wave (recall that $n=\int \frac{dp}{2\pi}f$ \cite{Groot1980RelativisticKT}),
\begin{equation}\label{densitone}
n = e^{i k x - k^{2} t} \int_{\mathbb{R}} \frac{dp}{2\pi} \, e^{-\beta\varepsilon - i k \beta p} \, ,
\end{equation}
is an exact solution of \eqref{diffusion}. By forming continuous superpositions of these modes, one may therefore generate a broad class of solutions to the diffusion equation. This mathematical result is not in conflict with causality. In fact, it is possible for a causal microscopic theory to contain an acausal hydrodynamic equation as an exact subsector \cite{GavassinoDiffusionCompatible2026tvy,GavassinoAcausalEmerge:2026fil}.

There is, however, an important caveat. For a kinetic state to be physically admissible (and for $n$ to be defined), the momentum integral in \eqref{densitone} must converge. In a Newtonian setting, where $\varepsilon = p^{2}/(2m)$, this integral is Gaussian and converges for all $k$ (since $\beta>0$). In the relativistic case, the situation is qualitatively different: at large momenta, the energy grows linearly, $\varepsilon(p) \sim |p|$. As a consequence, convergence requires that
\begin{equation}\label{thebound}
\boxed{|\mathfrak{Im}\, k| < 1 \, .}
\end{equation}
This condition constitutes the fundamental consistency requirement imposed by the underlying kinetic theory.\footnote{In \cite{GavassinoFokkerPlanck2026zsz}, an additional requirement was imposed, namely that the information current \cite{GavassinoCausality2021} be finite (equivalently, that $f$ belong to the Hilbert space defined by Onsager’s inner product \cite{GavassinoDistrubingMoving:2026klp}), leading to the more restrictive bound $|\mathfrak{Im}\, k|\leq 1/2$. In the present work, we only require finiteness of the density $n$, although the analysis can be straightforwardly adapted to incorporate stronger conditions.}

The bound \eqref{thebound} automatically ensures that the analysis is restricted to a class of modes for which diffusion is covariantly stable, in the sense that $\mathfrak{Im}\,\omega \leq |\mathfrak{Im}\, k|$ \cite{HellerBounds2022ejw,GavassinoBounds2023myj}. Indeed, since $\mathfrak{Im}\,\omega = \mathfrak{Im}(- i k^{2}) = (\mathfrak{Im}\, k)^{2} - (\mathfrak{Re}\, k)^{2}$, one has $(\mathfrak{Im}\, k)^{2} - (\mathfrak{Re}\, k)^{2} \leq (\mathfrak{Im}\, k)^{2} \leq |\mathfrak{Im}\, k|$. Nevertheless, it is important to stress that the condition \eqref{thebound} is stronger than the mere requirement of covariant stability. For instance, the mode with $k = 3 + 2 i$ is excluded by equation \eqref{thebound}, while it still satisfies the covariant stability condition, since $\mathfrak{Im}\,\omega = -5 < |\mathfrak{Im}\, k| = 2$. The bound \eqref{thebound} therefore represents a genuine physical admissibility criterion, rather than a mere reformulation of stability for equation \eqref{diffusion}.

\subsection{Wavenumber bounds on boosted diffusion}
\vspace{-0.2cm}

Consider a density profile $n(t,x)$ constructed as a linear superposition of the modes \eqref{densitone}, and suppose that it can be written as a spatially localized perturbation $\delta n(t,x)$ around a homogeneous background $n_0$. If the perturbation decays sufficiently rapidly at spatial infinity and is sufficiently regular at $t=0$, then it belongs to a standard class of localizable functions (e.g. the Schwartz class, or $L^2$, or $H^s$). In this case, $\delta n(t,x)$ admits a representation as a continuous superposition of \textit{Fourier modes}, namely of excitations of the form \eqref{densitone} with $k\in\mathbb{R}$. All such modes automatically satisfy the bound \eqref{thebound}. Consequently, aside from the usual convergence requirements associated with Fourier integrals \cite{GavassinoDiffusionCompatible2026tvy}, this construction allows one to span all localized smooth initial conditions $\delta n(0,x)$.

\begin{figure}[b!]
    \centering
\includegraphics[width=0.43\linewidth]{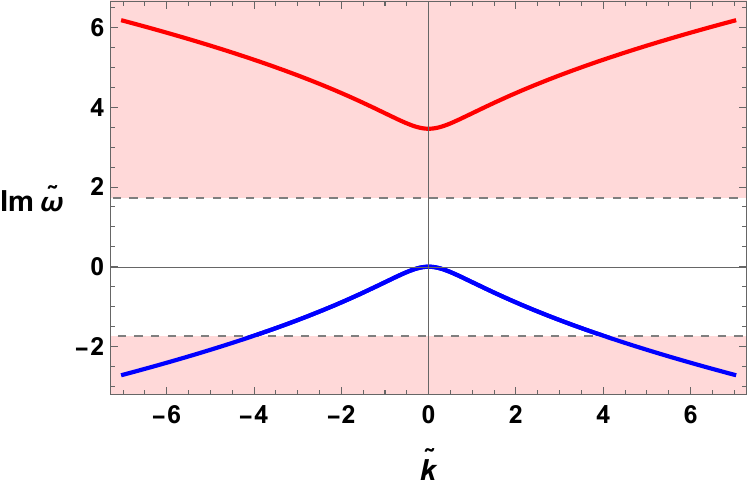}\hspace{0.08\linewidth}
\includegraphics[width=0.43\linewidth]{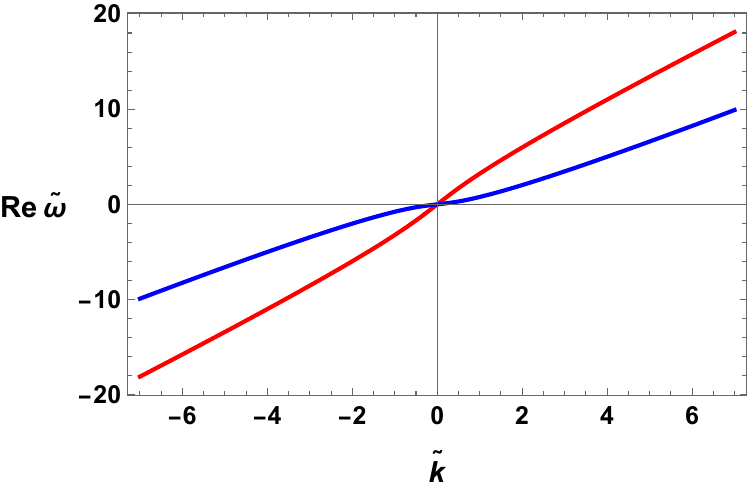}
\caption{Imaginary part (left) and real part (right) of the boosted dispersion relations
$\Tom(\Tk):\mathbb{R}\to\mathbb{C}$, obtained by applying a Lorentz boost to Fick’s law of diffusion
$\omega=-ik^{2}$. The boost velocity is chosen as $v=1/2$, but the qualitative features are
the same for all $v>0$. The blue branch is stable, while the red branch is unstable. The shaded region in the left panel indicates the kinetically
inadmissible domain, which must be excluded on physical grounds (the dashed lines correspond to $\mathfrak{Im}\, \omega=\pm 1/(\gamma v)$). As expected, the unstable branch
lies entirely within this region. In addition, the kinetic bound excludes the high-wavenumber portion
of the stable branch, leading to an upper cutoff in $\Tk$ (for $v=1/2$, the cutoff wavenumber is $\Lambda=4$).}
    \label{fig:Cutoff}
\end{figure}

\begin{figure}[h!]
    \centering
\includegraphics[width=0.43\linewidth]{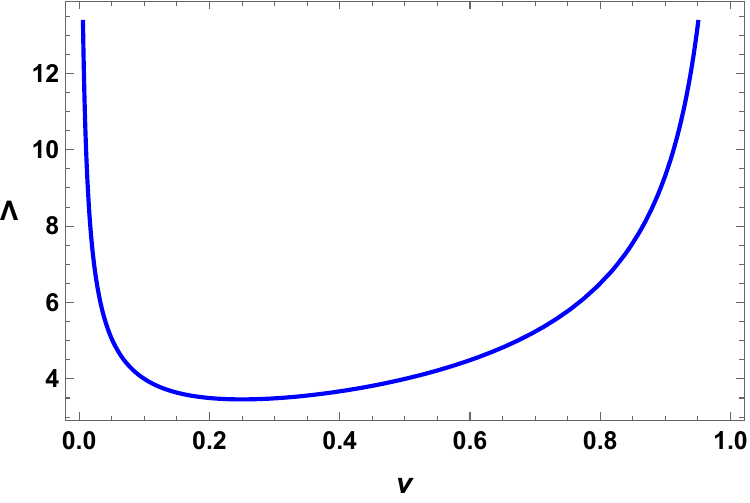}
\caption{Wavenumber cutoff $\Lambda$ as a function of the boost velocity $v$, as determined by
\eqref{cutoffone}. For $|\Tk|>\Lambda$, solutions of the boosted diffusion equation
\eqref{diffusionBoosted} no longer admit a realization within the underlying Fokker-Planck
kinetic theory, because the kinetic integral in \eqref{densitone} diverges.}
    \label{fig:CutoffScale}
\end{figure}

Let us now repeat the same reasoning in a reference frame in which the background medium moves with velocity $v\,{>}\,0$. We again assume that, in this boosted frame, the density perturbation $\delta n(\Tt,\Tx)$ is well defined everywhere at $\Tt=0$, so that a meaningful initial-value problem can be posed in this frame. We also assume that the initial state $\delta n(0,\Tx)$  is spatially localized and sufficiently regular, so that it may be expanded in modes that are of Fourier type \emph{in the boosted frame}, namely excitations of the form \eqref{densitone} with $\Tk\,{\in}\,\mathbb{R}$ \cite{Kost2000}. Here, the boosted wavevector $(\Tom,\Tk)$ is related to the original wavevector $(\omega,k)$ appearing in the ansatz \eqref{ansatzsolution} via $e^{i\Tk\Tx-i\Tom\Tt}\,{\equiv}\, e^{ikx-i\omega t}$, giving
\vspace{-0.1cm}
\begin{equation}\label{boostwavemode}
\begin{cases}
\omega=\gamma(\Tom-v\Tk)\, ,\\
k=\gamma(\Tk-v\Tom)\, .
\end{cases}
\end{equation}
In contrast to the unboosted case, however, the kinetic admissibility condition \eqref{thebound} now imposes nontrivial restrictions on the set of allowed Fourier modes. In fact, substituting the second relation in \eqref{boostwavemode} into \eqref{thebound}, and using the assumptions $\Tk\in\mathbb{R}$ and $v>0$, one finds the constraint
\vspace{-0.2cm}
\begin{equation}\label{secondbound}
|\mathfrak{Im}\,\Tom| < \dfrac{1}{\gamma v}\, .
\end{equation}
This restriction has two immediate and physically significant consequences, illustrated in
figure \ref{fig:Cutoff}. First, all exponentially growing solutions of the boosted diffusion
equation \eqref{diffusionBoosted} (such as $e^{\Tt/(\gamma v^{2})}$) are ruled out \textit{a priori}. This follows directly from the kinetic
admissibility condition \eqref{thebound}, which enforces covariant stability of the allowed
solutions.
Second, there is a wavenumber cutoff in the stable branch, meaning that all modes with $|\Tk|\geq \Lambda$ are excluded. To determine the explicit value of the cutoff scale $\Lambda$, one substitutes the boost
relations \eqref{boostwavemode} into the diffusion dispersion relation $\omega=-ik^{2}$, and
solves for $\Tk$ under the saturation condition $\mathfrak{Im}\,\Tom=-1/(\gamma v)$. This yields (see Appendix \ref{BBB})
\begin{equation}\label{cutoffone}
\Lambda=\dfrac{1+2v}{\sqrt{v(1-v)}} \, .
\end{equation}
As expected, the cutoff $\Lambda$ diverges in the limit $v\to 0$, reflecting the fact that, in the
rest frame of the medium, all wavenumbers are physically admissible. Interestingly, $\Lambda$
also diverges as $v\to 1$. As a result, the cutoff scale exhibits a minimum at $v=1/4$ (see
figure \ref{fig:CutoffScale}), where $\Lambda=2\sqrt{3}\approx 3.464$.

\subsection{Boosted initial value problem and well-posedness}
\label{sec:BoostedIVP}

The above analysis allows us to single out a natural class of \textit{physically admissible} initial data for moving observers. The boosted diffusion equation \eqref{diffusionBoosted} is thus endowed with the following physically-motivated initial-value formulation.

First, localized perturbations $\delta n$ at $\Tt=0$ are required to belong to a space of band-limited functions. 
Concretely, the initial profile is assumed to admit a Fourier representation of the form
\begin{equation}\label{initialData}
\delta n(0,\Tx)
=
\int_{-\Lambda}^{\Lambda}
\frac{d\Tk}{2\pi}\,
\varphi(\Tk)\,
e^{i\Tk\Tx},
\qquad 
\varphi \in L^2([-\Lambda,\Lambda]) \, ,
\end{equation}
where the cutoff wavenumber $\Lambda$ is given by \eqref{cutoffone}, and enforces the kinetic admissibility condition. 

Second, the time evolution of each Fourier mode should be governed by the \emph{stable} dispersion relation
\begin{equation}\label{stableDispersion}
\Tom(\Tk)
=
\frac{\Tk}{v}
+
\frac{i}{2\gamma v^2}
\left(
1-\sqrt{1-\frac{4iv\Tk}{\gamma}}
\right),
\end{equation}
which corresponds to the blue line in figure \ref{fig:Cutoff}.

Under these assumptions, the following properties hold.
\begin{itemize}
\item
Since $L^2([-\Lambda,\Lambda])\subset L^1([-\Lambda,\Lambda])$, the Fourier integral in \eqref{initialData} converges absolutely. 
As a consequence, the initial profile $\delta n(0,\Tx)$ is well-defined pointwise for all $\Tx\in\mathbb R$.

\item
The set of admissible initial data defined by \eqref{initialData} coincides with the Paley-Wiener space $PW_{\Lambda}$ \cite{PaleyWiener1934FourierTransforms,Zayed2008Bandlimitedness}, namely the closed subspace of $L^2(\mathbb R)$ consisting of functions whose Fourier transform is supported in $[-\Lambda,\Lambda]$. 
In particular, $PW_{\Lambda}$ is a Hilbert space when endowed with the $L^2$ norm.

\item
For every fixed $\Tt\in\mathbb R$, the evolution factor $e^{-i\Tom(\Tk)\Tt}$ satisfies
\(|e^{-i\Tom(\Tk)\Tt}|\leq e^{|\Tt|/(\gamma v)}\).
It follows that the evolved profile $\delta n(\Tt,\Tx)$ is well defined for all times, remains in $PW_{\Lambda}$, and satisfies the estimate
\begin{equation}\label{L2bound}
\|\delta n(\Tt)\|_{L^2}
\le
e^{|\Tt|/(\gamma v)}
\,
\|\delta n(0)\|_{L^2}.
\end{equation}
\end{itemize}

Putting everything together, we conclude that the boosted diffusion equation \eqref{diffusionBoosted}, when restricted to such kinetically admissible initial data and evolved along the stable branch, defines a well-posed initial value problem \cite[\S 3.10]{rauch2012partial}. 
The dynamics depends continuously on the initial data, and is well defined both forward and backward in time.

Let us also remark that applying any differential operator 
${\partial_{\Tx}}^a{\partial_{\Tt}}^b$ ($a,b\in\mathbb N$) to the Fourier
representation of $\delta n$ amounts to multiplying the integrand by
$(i\Tk)^a(-i\Tom)^b$. But since $\Tk$ and $\Tom(\Tk)$ are both bounded on the interval $[-\Lambda,\Lambda]$, the resulting Fourier coefficients still
belong to $L^2([-\Lambda,\Lambda])$. It follows that all
derivatives can be taken under the integral sign, and that
$\delta n\in C^\infty(\mathbb R^2)$. In particular,
$\delta n(\Tt,\Tx)$ solves the boosted diffusion equation \eqref{diffusionBoosted}
in the ordinary (i.e. non-distributional) sense globally.

\newpage
\section{General properties of the initial data}
\vspace{-0.3cm}

In the previous section, we showed that kinetic admissibility restricts
localized perturbations to the Paley-Wiener space $PW_{\Lambda}$ at all
times. This is a strong constraint, which severely limits the class of
admissible initial data \cite{Zayed2008Bandlimitedness}. In this section, we briefly examine its main
consequences.

\vspace{-0.4cm}
\subsection{Regularity}\label{regularity}
\vspace{-0.3cm}

Consider again the representation \eqref{initialData}, and expand the exponential
$e^{i\Tk\Tx}$ in powers of $\Tk\Tx$. Assuming that the resulting series may
be interchanged with the integral, one obtains
\vspace{-0.1cm}
\begin{equation}\label{itisentire}
\delta n(0,\Tx)
=
\sum_{a=0}^{\infty}\frac{(i\Tx)^a}{a!}
\int_{-\Lambda}^{\Lambda}
\frac{d\Tk}{2\pi}\,
\varphi(\Tk)\,\Tk^a .
\end{equation}
Since the magnitude of the integrals appearing in \eqref{itisentire} is bounded by
$\frac{1}{2\pi}\|\varphi\|_{L^1}\Lambda^a$, the series converges absolutely. This
justifies the exchange of summation and integration. It follows that
$\delta n(0,\Tx)$ is not only $C^\infty$, but can be analytically continued to an
entire function on the complex plane. Hence, it inherits all the standard properties of real-analytic functions. For example, it cannot be
compactly supported: if it vanishes on an interval, it must vanish
everywhere.

To illustrate the nontrivial implications of this result, consider the following
example. Let $G(t,x)$ denote the retarded Green function of the (unboosted)
diffusion equation \eqref{diffusion}, defined by
$(\partial_t-\partial_x^2)G=\delta(t)\delta(x)$ and $G(0^-,x)=0$, with explicit form
$G(t,x)=e^{-x^2/(4t)}/\sqrt{4\pi t}$ \cite[\S 7.4]{MorseFeshbach1953}. Applying a Lorentz boost yields \cite{GavassinoSuperlum2021}
\vspace{-0.1cm}
\begin{equation}\label{Gtx}
G(\Tt,\Tx)
=
\frac{\Theta(\Tt-v\Tx)}{\sqrt{4\pi\gamma(\Tt-v\Tx)}}
\exp\!\left[
-\frac{\gamma(\Tx-v\Tt)^2}{4(\Tt-v\Tx)}
\right].
\end{equation}
This function solves the boosted diffusion equation \eqref{diffusionBoosted}
everywhere except at the point $(\Tt,\Tx)=(0,0)$ \cite[\S 1.7, Probl.3]{rauch2012partial}. In particular, if we restrict
attention to the region $\Tt\ge1$, it provides an exact solution of
\eqref{diffusionBoosted} that is smooth (indeed~$C^\infty$), rapidly decaying at
spatial infinity, and supported exclusively on the stable branch of
modes\footnote{One can show (see Appendix \ref{ccc}) that the portion of $G$ with $\Tt<0$ is a
superposition of unstable modes (red curve in Fig.~\ref{fig:Cutoff}), whereas for
$\Tt>0$ it involves only stable modes (blue curve in Fig.~\ref{fig:Cutoff}).}. At
first sight, this might appear to define a perfectly admissible physical solution
of our initial-value problem, but that is not the case. The function \eqref{Gtx} vanishes identically in the region
$\Tx>\Tt/v$, and therefore cannot be entire in $\Tx$. As a consequence, it does not belong
to the Paley-Wiener space $PW_{\Lambda}$, meaning that its Fourier transform
contains contributions from arbitrarily large $|\Tk|$, including modes with
$|\Tk|>\Lambda$ (as verified in Appendix \ref{ccc}). A simple way to see this is to evolve the solution backward in
time: as $\Tt\to0^+$, a singularity is encountered at $\Tx=0$, demonstrating that
$G(1,\Tx)$ does not belong to a function space for which the initial-value problem
is well posed backward in time.

Despite the above mathematical discussion, it still may not be entirely obvious (from a physics perspective) why the restriction of $G$ to $\Tt\ge1$ cannot consistently arise as the
density associated with a solution of Fokker-Planck kinetic theory. Below, we show that the obstruction has a direct physical origin, which can be traced back to causality at the level of the kinetic equation.

Let us interpret the retarded Green function $G(t,x)$ as describing
the following process: for $t<0$ the system is in equilibrium, at
$t=0$ a unit of charge is injected at $x=0$, and for $t>0$ the excess
density spreads diffusively. In kinetic-theory language, this
corresponds to an equilibrium distribution function $f$ for $t<0$,
followed by the insertion of a perturbation at $t=0$, which then
relaxes under the Fokker-Planck dynamics. Now, the key subtlety is that, since the relativistic Fokker-Planck equation \eqref{fokkerplanck} is causal, if a kinetic perturbation $\delta f$ is truly localized at $x=0$, it cannot propagate outside the lightcone. By contrast, $G(t,x)$ exhibits infinitely long spatial tails for all $x$ at any $t>0$. This implies that, even though the density perturbation $\delta n$ is localized at $x=0$ at $t=0$, the corresponding perturbation $\delta f$ must have been generated by a kinetic source that was already nonlocal at that time\footnote{Indeed, it is shown in \cite{GavassinoDiffusionCompatible2026tvy} that no perturbation $\delta f$ belonging to the diffusive sector can be compactly supported, even when the associated density perturbation $\delta n$ is. Hence, the apparent acausal spreading of the density is reconciled with microcausality through the fact that the hydrodynamic projection is spatially nonlocal within the underlying kinetic theory.}.
In other words, while the diffusion Green function $G$ is generated by a source term $\delta(t)\delta(x)$ at the level of the diffusion equation, generating $G$ as the exact density associated with a solution of Fokker-Planck (without additional non-hydrodynamic contributions) requires a more complicated source term $\delta(t)\sigma(x,p)$ that is infinitely extended in space, i.e. $\sigma(x,p)\not\propto\delta(x)$.

As a consequence, when one performs a Lorentz boost and restricts
attention to the region $\Tt\ge1$, the resulting density profile
solves the boosted diffusion equation \eqref{diffusionBoosted}
everywhere in that region. However, its associated distribution
function fails to satisfy the kinetic equation \eqref{fokkerplanck}
along the line $\Tx=\Tt/v$, where a source term appears. This line coincides with the boundary of
the region where $G$ vanishes identically, and therefore marks the
precise locus at which analyticity, and hence kinetic admissibility,
breaks down.

\subsection{Bounds on localizability}

The impossibility for $\delta n(0,\Tx)$ to be compactly supported is only
one manifestation of the strong localization constraints obeyed by
Paley-Wiener functions. Additional limitations follow from standard
uncertainty-type arguments. In particular, the uncertainty relation
$\Delta \Tx\,\Delta \Tk \ge 1/2$ implies a lower bound on spatial
localization. Since functions in $PW_{\Lambda}$ have Fourier support of
length $2\Lambda$, one has $\Delta \Tk \le 2\Lambda$, and therefore
\begin{equation}
\Delta \Tx > (4\Lambda)^{-1}.
\end{equation}
The inequality is strict, as saturation of the uncertainty bound occurs
only for Gaussian wavepackets, which do not belong to the Paley-Wiener
class.
A related constraint follows from a pointwise bound on the density
profile. Indeed, from $|\delta n(\Tx)|\le \frac{1}{2\pi}
\|\varphi\|_{L^1}=\frac{1}{2\pi}(e^{i\arg\varphi},\varphi)$ and the
Cauchy-Schwarz inequality, one finds \cite{Brevig2022}
\begin{equation}
|\delta n(\Tx)|\le
\sqrt{\dfrac{\Lambda}{\pi}}\,\|\delta n\|_{L^2}.
\end{equation}
Both bounds indicate that there is a fundamental limit to how strongly
the density can be concentrated at a single location. If either bound
is violated at a given time, a boosted observer can immediately infer
that the subsequent density evolution cannot be governed solely by the
diffusive sector of Fokker-Planck kinetic theory, but must involve
non-hydrodynamic contributions. In fact, in addition to the diffusive mode, Fokker-Planck also contains two quasi-ballistic branches \cite{GavassinoFokkerPlanck2026zsz}, into which the diffusive mode is absorbed when \eqref{thebound} is violated. Such modes inevitably appear whenever one attempts to localize the boosted kinetic initial data over lengthscales shorter than $\Lambda^{-1}$.

\subsection{Shannon-Whittaker representation}

The final structural property of the space $PW_{\Lambda}$ that will be
relevant for our purposes is the Shannon-Whittaker sampling theorem \cite{Garcia2009BriefWalkSampling},
whose content we briefly recall below.

The starting point is the observation that the set
$\{e^{-i\pi a \Tk/\Lambda}\}_{a\in\mathbb Z}$ forms an orthogonal basis
of $L^2([-\Lambda,\Lambda])$. As a consequence, the Fourier
representation \eqref{initialData} admits the decomposition
\begin{equation}
\delta n(0,\Tx)
=
\int_{-\Lambda}^{\Lambda}\frac{d\Tk}{2\pi}
\sum_{a=-\infty}^{+\infty} c_a\,
e^{i\Tk(\Tx-\frac{\pi a}{\Lambda})}
=
\sum_{a=-\infty}^{+\infty} c_a
\int_{-\Lambda}^{\Lambda}\frac{d\Tk}{2\pi}\,
e^{i\Tk(\Tx-\frac{\pi a}{\Lambda})},
\end{equation}
which yields
\begin{equation}
\delta n(0,\Tx)
=
\sum_{a=-\infty}^{+\infty} c_a\,
\frac{\sin[\Lambda(\Tx-\frac{\pi a}{\Lambda})]}
{\pi(\Tx-\frac{\pi a}{\Lambda})}.
\end{equation}
Evaluating this expression at the sampling points
$\Tx_a=\pi a/\Lambda$ immediately gives
$c_a=\pi\,\delta n(0,\Tx_a)/\Lambda$, and therefore
\begin{equation}
\delta n(0,\Tx)
=
\sum_{a=-\infty}^{+\infty}
\delta n(0,\Tx_a)\,
\frac{\sin[\Lambda(\Tx-\Tx_a)]}{\Lambda(\Tx-\Tx_a)}.
\end{equation}
This identity shows that any initial profile $\delta n(0,\Tx)\in
PW_{\Lambda}$ can be uniquely reconstructed from its values on the
discrete set of points $\Tx_a=\pi a/\Lambda$, which motivates the
terminology ``sampling theorem''.

An immediate consequence is that, once the time evolution of the
initial profile $\mathrm{sinc}(\Lambda\Tx)$ is known, the corresponding
solution $\mathcal{K}(\Tt,\Tx)$ can be used to generate all other
solutions via the expansion
\begin{equation}\label{samplingformula}
\delta n(\Tt,\Tx)
=
\sum_{a=-\infty}^{+\infty}
\delta n(0,\Tx_a)\,
\mathcal{K}(\Tt,\Tx-\Tx_a).
\end{equation}
In this sense, $\mathcal{K}$ plays the role of a ``discrete Green''
function, or fundamental solution, for boosted diffusion.

It is worth stressing, however, that $\mathcal{K}$ is not a
Green function in the usual sense (and it is not retarded). In fact, since $\mathrm{sinc}(\Lambda\Tx){\in}
PW_{\Lambda}$, the function $\mathcal{K}(\Tt,\Tx)$ is necessarily
smooth and real-analytic, and therefore cannot vanish on any dense set
of spacetime points. Accordingly, $\mathcal{K}$ solves the boosted
diffusion equation \eqref{diffusionBoosted} throughout $\mathbb R^2$,
\textit{without} the appearance of Dirac-delta source terms at $(\Tt,\Tx)=(0,0)$, and it describes
both forward and backward time evolution.

We also emphasize that the discrete representation \eqref{samplingformula} is not an
approximation scheme, nor a numerical truncation. Rather, it is an
exact identity within the space $PW_{\Lambda}$. The emergence of a
discrete sampling structure is the direct counterpart of the strong
limitations on spatial localization imposed by band limitation, which ultimately originates from the bound \eqref{thebound}.

\section{The fundamental solution}

In this section, we compute the function $\mathcal{K}(\Tt,\Tx)$
explicitly, and analyze its properties. This function provides the
basic building block from which any other kinetically admissible
solution of \eqref{diffusionBoosted} can be reconstructed, through the
sampling formula \eqref{samplingformula}.

\subsection{Expressing the Fourier integral in the rest frame}

Since the initial condition is $\mathcal{K}(0,\tilde{x})=\mathrm{sinc}(\Lambda\tilde{x})$, the fundamental solution admits the integral representation
\begin{equation}
\mathcal{K}(\tilde{t},\tilde{x})
=\int_{-\Lambda}^{\Lambda}\frac{d\tilde{k}}{2\Lambda}\,
e^{i\tilde{k}\tilde{x}-i\tilde{\omega}\tilde{t}},
\end{equation}
where $\tilde{\omega}(\tilde{k})$ is the stable dispersion relation, given in \eqref{stableDispersion}. In this form, there is no hope of evaluating the integral explicitly, since $\tilde{\omega}$ appears in the exponent as a complicated function of $\tilde{k}$. However, we can transform back to the rest frame, where the dynamics are much simpler.

By construction, $e^{i\tilde{k}\tilde{x}-i\tilde{\omega}\tilde{t}}=e^{ikx-k^2 t}$. Accordingly, the integral over $\tilde{k}\in[-\Lambda,\Lambda]$ may be rewritten as an integral over the complex $k$-plane along the image of this interval under the mapping $k(\tilde{k})=\gamma(\tilde{k}-v\tilde{\omega}(\tilde{k}))$. Since $k(\tilde{k})$ is analytic and one-to-one on $[-\Lambda,\Lambda]$, this change of variables defines a smooth contour in the complex $k$-plane with endpoints $k(-\Lambda)$ and $k(\Lambda)$. This contour avoids the branch singularities associated with the square root in $\tilde{\omega}$. For the cutoff value $\Lambda$ used here, the endpoints take the simple form (see Appendix~\ref{BBB})
\begin{equation}\label{wavevenumbercutoffs}
k(\pm\Lambda)= i \pm \sqrt{1+\frac{1}{v}} \, .
\end{equation}
Moreover, the relation $k(\tilde{k})$ can be inverted to yield $\tilde{k}=\gamma[k+v\omega(k)]=\gamma(k-i v k^2)$, so that the measure transforms as $d\tilde{k}=\gamma(1-2ivk)\,dk$. We therefore obtain
\begin{equation}\label{restnonfourier}
\mathcal{K}(t,x)
=\int_{k(-\Lambda)}^{k(\Lambda)}
\frac{dk}{2\Lambda}\,
\gamma(1-2ivk)\,e^{ikx-k^2 t}.
\end{equation}
Since the integrand is an entire function of $k$, the value of the integral depends only on the endpoints of the contour (and not on the particular choice of path connecting them).

Before evaluating the integral explicitly for generic values of $t$
and $x$, it is instructive to verify that \eqref{restnonfourier}
indeed possesses the properties required of $\mathcal{K}$. First,
by construction, it is an exact solution of the diffusion equation, being a superposition of modes $\propto e^{ikx-k^2 t}$.
Moreover, the integral converges for all $(t,x)\in\mathbb{R}^2$ and
defines a real-analytic function, by the same arguments discussed in
Section~\ref{regularity}. The integration contour may be chosen as
\begin{equation}
k=\mathfrak{Re}\,k+i\,\frac{(\mathfrak{Re}\,k)^2}{1+1/v},
\qquad\qquad
\mathfrak{Re}\,k\in\left(
-\sqrt{1+\frac{1}{v}},\,
\sqrt{1+\frac{1}{v}}
\right),
\end{equation}
which lies entirely within the strip \eqref{thebound} and is therefore
compatible with kinetic admissibility.

The final property to be checked is the initial condition. When
$\Tt=0$, the expression \eqref{restnonfourier} must reduce to
$\mathrm{sinc}(\Lambda\Tx)$. When $\Tt=\gamma(t+vx)=0$,
the prefactor appearing in the integrand becomes proportional to the
derivative of the exponential, allowing the integral to be evaluated
explicitly. One finds
\begin{equation}\label{restnonfourier2}
\begin{split}
\mathcal{K}(-vx,x)
={}&
\int_{k(-\Lambda)}^{k(\Lambda)}
\frac{dk}{2\Lambda}\,
\gamma(1-2ivk)\,e^{ikx+vk^2x}
=
\frac{\gamma}{2i\Lambda x}
\int_{k(-\Lambda)}^{k(\Lambda)}
dk\,
(ix+2vkx)\,e^{ikx+vk^2x}
\\
={}&
\frac{\gamma}{2i\Lambda x}
\left(
e^{ik(\Lambda)x+vk(\Lambda)^2x}
-
e^{ik(-\Lambda)x+vk(-\Lambda)^2x}
\right)
\\
={}&
\frac{\gamma}{\Lambda x}
\sin\!\left(
\frac{x}{\gamma}\,
\frac{1+2v}{\sqrt{v(1-v)}}
\right)
=
\mathrm{sinc}(\Lambda\Tx),
\end{split}
\end{equation}
as required.

\subsection{Evaluating the Fourier integral}

\begin{figure}[b!]
    \centering
\includegraphics[width=0.43\linewidth]{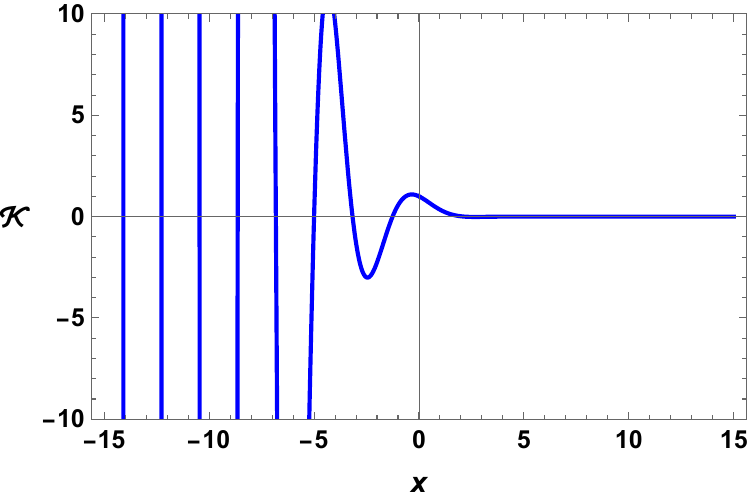}
\caption{Snapshot at $t=0$, in the rest frame of the medium, of the fundamental
solution $\mathcal{K}$ associated with the boosted initial-value
problem for an observer moving with velocity $-1/2$. The profile
is modulated by an exponential factor $e^{-x}$, which arises due to relativity of simultaneity, and reflects the saturation of the kinetic admissibility bound \eqref{thebound}. The corresponding analytic formula is
given in equation \eqref{iniziamofund}.
}
    \label{fig:Initialdatafund}
\end{figure}

Since \eqref{restnonfourier} is a contour integral of an entire
function, its value is given by the difference of the
primitive evaluated at the endpoints of the contour.
Recalling that derivatives may be taken outside the integral sign, one
finds
\begin{equation}\label{TheAnswer} 
\mathcal{K}(t,x) =\dfrac{\gamma \sqrt{\pi }}{4\Lambda \sqrt{t}} (1-2v \partial_x) \left[ e^{-\frac{x^2}{4 t}}\text{erf}\left(\sqrt{t}\, k(\Lambda)-\dfrac{ix}{2\sqrt{t}} \right)- e^{-\frac{x^2}{4 t}}\text{erf}\left(\sqrt{t}\, k(-\Lambda)-\dfrac{ix}{2\sqrt{t}}\right)\right]\, . \end{equation}
This expression is well defined also for negative values of $t$, in
which case, the square root contributes a factor of $i$. Equivalently,
for $t<0$ one may replace $\sqrt{t}$ with $\sqrt{|t|}$ and the error
function $\text{erf}$ with the imaginary error function $\text{erfi}$. Moreover,
evaluating \eqref{restnonfourier} directly at $t=0$ yields (see figure \ref{fig:Initialdatafund})
\begin{equation}\label{iniziamofund}
\mathcal{K}(0,x) =\dfrac{1}{1{+}2v}(1-2v\partial_x)\left[e^{-x}\, \text{sinc}\left(x\sqrt{1+\dfrac{1}{v}} \right) \right]\, . 
\end{equation}

The appearance of the exponential factor $e^{-x}$ deserves some
explanation. In constructing $\mathcal{K}$, we are effectively asked
to determine an initial density profile $\mathcal{K}(0,x)$ such that,
upon time evolution, it reproduces the profile
$\text{sinc}(\Lambda\Tx)$ on the simultaneity line $\Tt=0$ (i.e. $t=-vx$) of an observer
$\mathcal{O}$ moving with velocity $-v$ (with $v>0$). For large
positive $x$, this simultaneity line lies in the far past relative to $t=0$, so one expects that diffusion has exponentially damped the
profile by the time it reaches the line $t=0$. Accordingly,
$\mathcal{K}(0,x)$ must decay exponentially for $x\to+\infty$. By contrast, for large negative $x$, the simultaneity
line $\Tt=0$ lies far in the future of $t=0$, so the initial profile must contain
oscillations of exponentially large amplitude in order for them to survive up to
$\Tt=0$. The fact that the resulting asymmetry is precisely encoded by
the factor $e^{-x}$ reflects the circumstance that $\mathcal{K}$ is
constructed as a superposition of all kinematically admissible modes
in the boosted frame, and therefore lies at the boundary of the
kinetic admissibility condition \eqref{thebound}.

From a mathematical standpoint, this behavior illustrates that the
natural class of spatially localized functions for which the boosted
initial-value problem is well posed cannot be obtained by simply
boosting solutions that are spatially localized in the rest frame. In this
sense, spatial localizability is a frame-dependent notion for
diffusive dynamics. This sharply contrasts with the situation in
causal theories such as Cattaneo's, where compact support in one reference
frame implies compact support in all reference frames \cite{GavassinoSuperlum2021}.

Finally, it is instructive to consider the limits of small and large
boost velocities. As $v\to0$, one finds that $\mathcal{K}(0,x)\approx \text{sinc}(x/\sqrt{v})$, which converges to $\pi \sqrt{v}\delta(x)$. This behavior is expected: in the
limit $v\to0$, the boosted initial-value problem reduces to the
rest-frame one, the cutoff $\Lambda$ diverges as $1/\sqrt{v}$, and the
sampling formula \eqref{samplingformula} becomes indistinguishable
from the usual Green-function representation, with the replacement
$\sum_a\to\int dx/(\pi\sqrt{v})$.
In the opposite limit $v\to1$, no singular behavior occurs, and
$\mathcal{K}$ converges to a finite and smooth function. This reflects
the fact that both the simultaneity hypersurface of $\mathcal{O}$ and
the cutoff wavenumbers \eqref{wavevenumbercutoffs} approach finite limits.
The fact that $\Lambda$ diverges as $v\rightarrow 1$ (see
figure \ref{fig:CutoffScale}) is entirely due to the
contraction of the boosted coordinates. There is no intrinsic
singularity in $\mathcal{K}$ itself.

\subsection{Building exact solutions}
\vspace{-0.2cm}

\begin{figure}[b!]
    \centering
\includegraphics[width=0.43\linewidth]{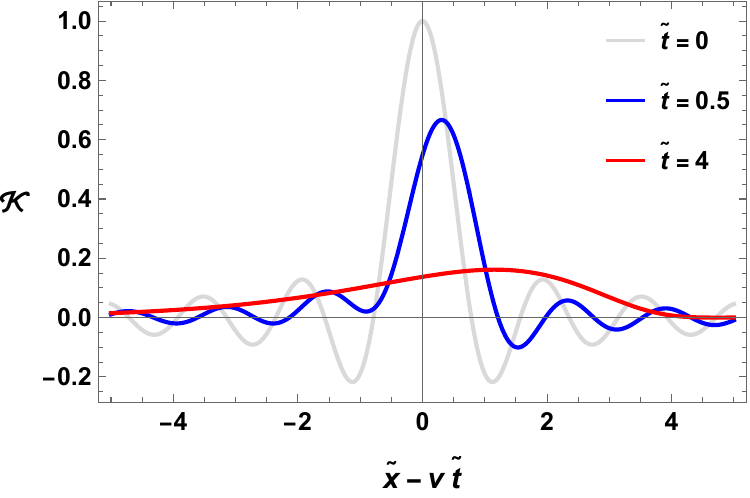}\hspace{0.08\linewidth}
\includegraphics[width=0.43\linewidth]{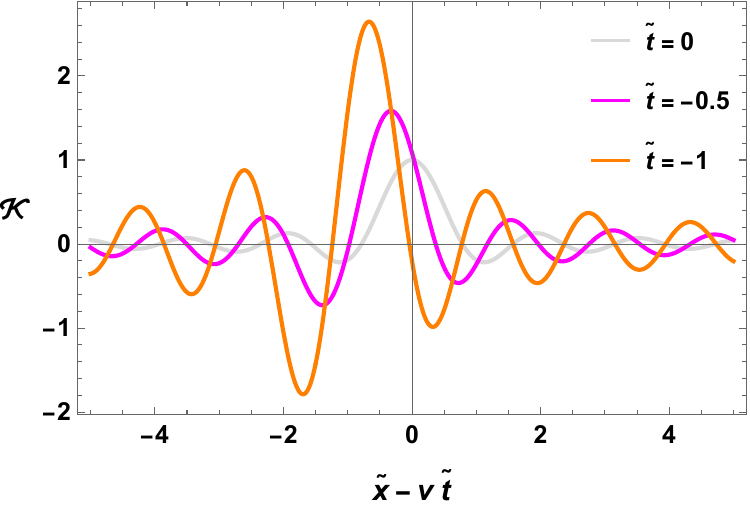}
\caption{
Snapshots of the forward (left panel) and backward (right panel) time evolution of the Green function $\mathcal{K}(\tilde t,\tilde x)$, obtained by boosting \eqref{TheAnswer}.
To compensate for advection by the moving medium, each snapshot covers a region shifted by $v\tilde t$.
For $\tilde t>0$, the evolution exhibits asymmetric diffusion and rapid suppression of oscillations associated with the cutoff scale.
Under backward time evolution, those same oscillatory features are progressively amplified. Again, we chose $v=1/2$.
}
    \label{fig:Fundamental}
\end{figure}

By boosting the expression \eqref{TheAnswer} back to the moving frame,
one obtains the Green function $\mathcal{K}(\Tt,\Tx)$ (see figure 
\ref{fig:Fundamental}), which enters the sampling formula
\eqref{samplingformula}. As such, $\mathcal{K}$ provides the basic
building block from which all solutions of the initial-value problem
introduced in Section~\ref{sec:BoostedIVP} can be reconstructed.

It is worth emphasizing that the sampling formula is most naturally
interpreted not only as an evolution rule, but also as a constructive
prescription for admissible initial data. Indeed, one may start from
an arbitrary sequence of coefficients
$\{c_a\}_{a\in\mathbb Z}\in\ell^2$ (which is needed for convergence), and
define
\begin{equation}\label{dnAfinal}
\delta n(\Tt,\Tx)
=
\sum_{a=-\infty}^{+\infty}
c_a\,
\mathcal{K}(\Tt,\Tx-\Tx_a),
\end{equation}
where $\Tx_a=\pi a/\Lambda$. By construction, the resulting profile
belongs to $PW_\Lambda$ and is thus kinetically admissible.
Moreover, it solves the boosted diffusion equation
\eqref{diffusionBoosted} exactly, and its initial profile satisfies
$\delta n(0,\Tx_a)=c_a$ at the sampling points.

Since $\mathcal{K}(0,\Tx)=\text{sinc}(\Lambda\Tx)$, the qualitative features
of the reconstructed initial profile are directly controlled by the
behavior of the coefficients $c_a$. If these vary rapidly from one
integer to the next, the resulting profile $\delta n(0,\Tx)$ will oscillate on the shortest
admissible spatial scale, of order $1/\Lambda$, which reflects the
presence of modes near the cutoff. Conversely, if the coefficients
are obtained by sampling a reference function $g(\Tx)$ that varies
smoothly on scales much larger than $1/\Lambda$, namely
$c_a=g(\Tx_a)$, then the reconstructed profile $\delta n(0,\Tx)$
provides an accurate $PW_\Lambda$-approximation of $g(\Tx)$. In this
regime, the effect of the cutoff becomes negligible, and the sampling
representation is effectively indistinguishable from a continuum
description.

\begin{figure}[h!]
    \centering
\includegraphics[width=0.43\linewidth]{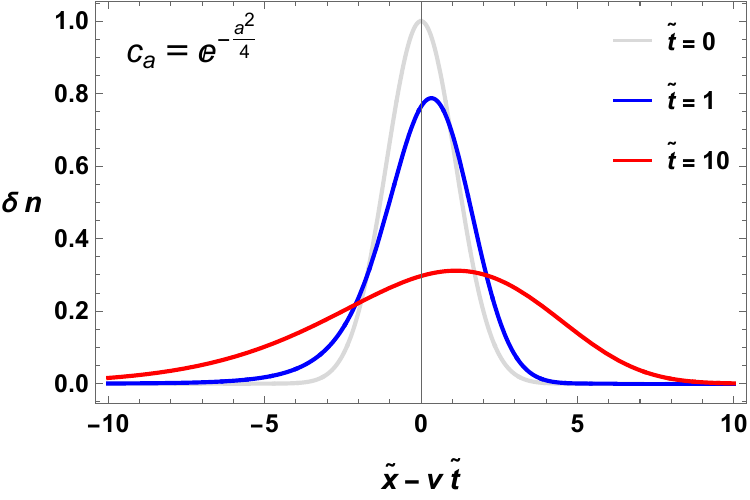}\hspace{0.08\linewidth}
\includegraphics[width=0.43\linewidth]{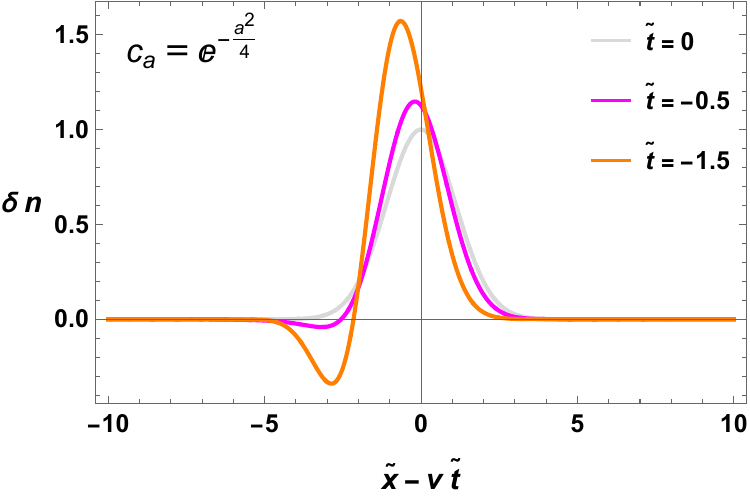}
\includegraphics[width=0.43\linewidth]{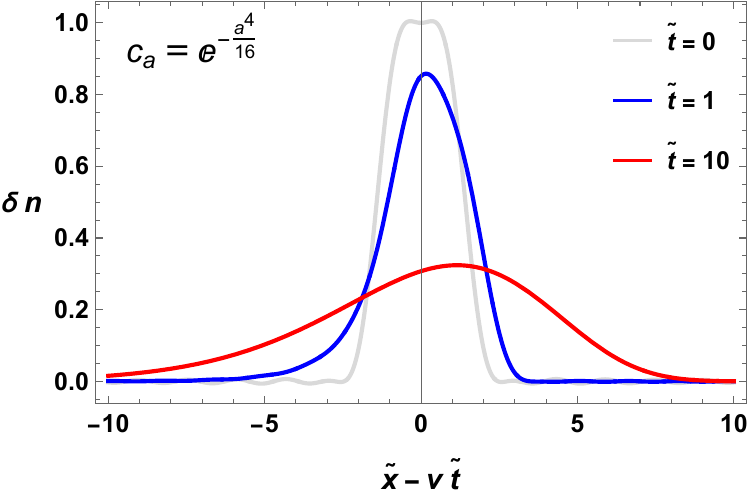}\hspace{0.08\linewidth}
\includegraphics[width=0.43\linewidth]{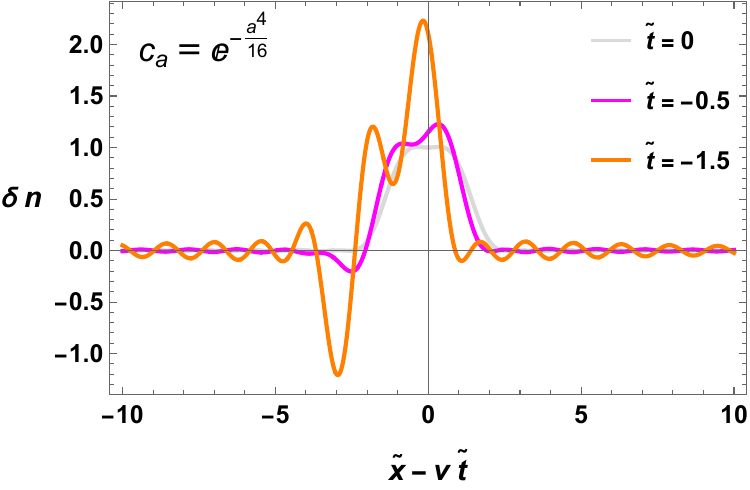}
\caption{ Exact solutions~\eqref{dnAfinal} of the boosted diffusion equation~\eqref{diffusionBoosted} for the initial data~\eqref{initialdataexampleapproximae} at boost velocity $v=1/2$.
Left panels: forward evolution. We see a progressive loss of sensitivity to the detailed structure of the initial profile, and convergence toward a universal (asymmetric) diffusive shape.
Right panels: backward evolution. Here, antidiffusion amplifies modes near the cutoff, leading to the emergence or enhancement of spatial oscillations on the scale $1/\Lambda$.
}
    \label{fig:ExactSolutions}
\end{figure}

As a simple illustration, one finds
\begin{equation}\label{initialdataexampleapproximae}
\vspace{-0.1cm}
\begin{split}
\sum_{a=-\infty}^{+\infty}
e^{-(\frac{a}{2})^2}\,
\mathcal{K}(0,\Tx-\Tx_a)
\approx{}&
e^{-(\frac{\Lambda\Tx}{2\pi})^2} \qquad (\text{very accurate})\, ,\\
\sum_{a=-\infty}^{+\infty}
e^{-(\frac{a}{2})^4}\,
\mathcal{K}(0,\Tx-\Tx_a)
\approx{}&
e^{-(\frac{\Lambda\Tx}{2\pi})^4} \qquad (\text{slightly worse, because more sharply localized}) \, .\\
\end{split}
\end{equation}
As can be seen in figure \ref{fig:ExactSolutions}, the oscillatory features associated with the cutoff scale are strongly suppressed\footnote{Note that, even if oscillations are not visible to the naked eye (at least in the case with $e^{-a^2/4}$), oscillatory tails are generically present for band-limited functions (although highly exceptional non-oscillating Paley-Wiener functions do exist~\cite{OstrovskiiUlanovskii2004NonOscillatingPW}). Such oscillatory components are progressively amplified under backward time evolution, as shown in figure \ref{fig:ExactSolutions}.
}, and they become progressively smaller at later times. 
Taken together, these examples indicate that, in the long-wavelength regime
(relative to the scale $1/\Lambda$),
and upon restricting attention to forward time evolution, the exact,
kinetically admissible solutions~\eqref{dnAfinal} are effectively indistinguishable
from those obtained by solving the boosted diffusion equation~\eqref{diffusionBoosted}
for generic smooth initial density profiles $n(0,\Tx)$, where the initial time derivative
$\partial_{\Tt} \, n(0,\Tx)$ is fixed so as to project out the unstable branch and retain
only the stable modes. In this regime, the kinetic admissibility constraint \eqref{thebound} plays no
dynamical role beyond enforcing stability forward in time.

\subsection{Kinetic embedding}

Up to this point, we have implicitly assumed that any solution of~\eqref{diffusion}
constructed as a linear superposition of kinetically admissible modes is itself
kinetically admissible. This assumption, however, hides a subtle mathematical issue.
Even if a superposition of density modes of the form~\eqref{densitone} converges to a
regular density profile $n(t,x)$, this does not automatically imply that the
corresponding superposition of kinetic modes~\eqref{ansatzsolution} also converges to a
regular distribution function $f(t,x,p)$ \cite{GavassinoDiffusionCompatible2026tvy}.
Therefore, in order to rigorously establish that the solutions~\eqref{dnAfinal} admit
a realization within Fokker-Planck kinetic theory, one must explicitly verify that
the kernel~\eqref{TheAnswer} arises as the particle density associated with a
physically admissible solution of the kinetic equation~\eqref{fokkerplanck}.

In the case of massless particles, it is known that, if such a kinetic realization
exists, it must take the form~\cite{GavassinoDiffusionCompatible2026tvy}
\begin{equation}\label{fofK}
\delta f_{\mathcal{K}}(t,x,p)
=\pi \beta e^{-\beta |p|}\,(1-\partial_x^2)\mathcal{K}(t,x-\beta p)\, .
\end{equation}
It is straightforward to verify that~\eqref{fofK} indeed solves
\eqref{fokkerplanck}. The associated density reads
\begin{equation}
\begin{split}
\delta n(t,x)
={}& \int_{\mathbb{R}} \frac{dp}{2\pi}\,\pi \beta e^{-\beta |p|}
(1-\partial_x^2)\mathcal{K}(t,x-\beta p)
\stackrel{\xi\equiv-\beta p}{=} \int_{\mathbb{R}} \frac{d\xi}{2}\, e^{-|\xi|}
(1-\partial_\xi^2)\mathcal{K}(t,x+\xi) \\
={}& \int_{-\infty}^0 \frac{d\xi}{2}\, e^{\xi}
(1-\partial_\xi^2)\mathcal{K}(t,x+\xi)
+ \int_{0}^{+\infty} \frac{d\xi}{2}\, e^{-\xi}
(1-\partial_\xi^2)\mathcal{K}(t,x+\xi) \\
={}& \mathcal{K}(t,x)
- \left[\frac{e^{\xi}}{2}\,(\mathcal{K}-\partial_\xi\mathcal{K})\right]_{\xi=-\infty}
- \left[\frac{e^{-\xi}}{2}\,(\mathcal{K}+\partial_\xi\mathcal{K})\right]_{\xi=+\infty} \, ,
\end{split}
\end{equation}
where the final line follows from integration by parts (recall that $\mathcal{K}$ is
smooth). The boundary term at $+\infty$ decays exponentially, as $e^{-2\xi}$, while
the contribution at $-\infty$ decays $1/\xi$. Hence, both these terms vanish at infinity, meaning that the integral converges
(and this is precisely our admissibility criterion), so~\eqref{fofK} defines a kinetically admissible solution of the Vlasov-Fokker-Planck equation
with associated density $\delta n(t,x)=\mathcal{K}(t,x)$, as required.

\section{Conclusions}
\vspace{-0.3cm}

In this work, we have reconsidered how an initial-value problem should be formulated for Lorentz-boosted diffusion, starting from its microscopic embedding within relativistic
Fokker-Planck kinetic theory \cite{GavassinoDiffusionCompatible2026tvy,GavassinoFokkerPlanck2026zsz}. By exploiting the fact that Fick-type diffusion arises as
the \emph{exact} hydrodynamic sector of the kinetic theory, we identified a physically
motivated space of admissible initial data for boosted profiles, based on the requirement that the resulting evolution should admit a kinetic realization. This criterion singles out a space of band-limited functions, and allows one to define an initial-value problem for the boosted diffusion equation that
is well posed and microscopically meaningful.

Within this restricted setting, we showed that the resulting dynamics are well defined not
only forward in time, but also backward. While the forward evolution undergoes ordinary diffusive smoothing,
the backward evolution is correspondingly ``anti-diffusive'', with spatial structures on the
shortest admissible scales being progressively amplified. Crucially, however, this growth
is not arbitrarily fast. The kinetic admissibility condition enforces a sharp wavelength cutoff,
which in turn bounds the growth rate to be controlled by a maximal exponent
$1/(\gamma v)$. As a result, the evolution remains well-posed in the sense of Hadamard \cite[\S 3.10]{rauch2012partial}: small
changes in the initial data lead to proportionally small changes in the solution at any
finite time. In physical terms, the cutoff prevents the catastrophic short-wavelength
instabilities that would otherwise render the backward problem ill-posed, and ensures a
controlled, continuous time-reversed dynamics within the kinetically admissible sector.

Finally, we have verified that the mathematical technicalities associated with band-limited function spaces become largely irrelevant in the regime that is most relevant for practical applications. In the long-wavelength limit, and upon restricting attention to forward time
evolution, the exact solutions constructed here are effectively indistinguishable from
those obtained by solving the boosted diffusion equation after simply discarding the
unstable branch of modes. From this perspective, the present analysis provides a
microscopic justification for what is otherwise the most obvious fix of boosted diffusion.

It is important to emphasize that the construction presented here is not intended as a
``practical'' framework for relativistic transport theory. In fact, our analysis is restricted to one
spatial dimension, to the linear regime, and to a homogeneous background medium. Indeed, it remains open whether an analogous kinetically admissible formulation can be extended beyond the linear regime, where nonlinear mode coupling may transfer energy toward progressively shorter wavelengths and potentially excite non-hydrodynamic degrees of freedom. Moreover, the whole construction relies heavily on the special structure of the Fokker-Planck kinetic equation. For these
reasons, we do not expect this approach to replace more conventional formulations of
relativistic dissipation \cite{cattaneo1958,Israel_Stewart_1979,BemficaDNDefinitivo2020}, nor to be directly applicable in fully realistic settings. Its primary
interest lies instead in its conceptual implications. Our results show that, in some cases, it is possible to make precise mathematical sense of evolution equations that
are acausal and linearly unstable when treated as stand-alone partial differential
equations, provided one is willing to impose non-local constraints
on the admissible initial data.

\vspace{-0.3cm}
\section*{Acknowledgements}
\vspace{-0.3cm}

This work is supported by a MERAC Foundation prize grant,  an Isaac Newton Trust Grant, and funding from the Cambridge Centre for Theoretical Cosmology.

\appendix

\vspace{-0.3cm}
\section{Cattaneo theory as an exact kinetic equation}\label{aaa}
\vspace{-0.3cm}

Following \cite{Basar:2024qxd}, we decompose the ensemble of massless particles into right movers, which propagate with velocity $+1$, and left movers, which propagate with velocity $-1$, disregarding the precise value of their momentum. In a kinetic-theory framework, where the fundamental degree of freedom is the single-particle distribution function $f(t,x,p)$, this amounts to introducing the partial densities
\begin{equation}
n_+(t,x)=\int_0^{+\infty} \frac{dp}{2\pi} f(t,x,p)  \, , \qquad \qquad  
n_-(t,x)=\int_{-\infty}^{0} \frac{dp}{2\pi} f(t,x,p) \, .
\end{equation}
Integrating the Boltzmann equation $(\partial_t + \V \partial_x)f=\text{``scattering integral''}$ (with $\V=\mathrm{sgn}(p)$) over positive or negative momenta, and invoking particle-number conservation, one obtains $(\partial_t+\partial_x)n_+=\mathcal{R}$ and $(\partial_t-\partial_x)n_-=-\mathcal{R}$. Here, $\mathcal{R}$ denotes the net rate of conversion between right movers and left movers induced by random scattering with the environment.
The central assumption of the model is that all particles, independently of their momentum, have the same probability per unit time of reversing their direction of motion. Under this assumption, the conversion rate takes the form $\mathcal{R}=(n_- - n_+)/2$ (in appropriate units). Introducing the total particle density $n=n_+ + n_-$ and the associated flux $J=n_+ - n_-$, one finds
\vspace{-0.3cm}
\begin{equation}
\begin{split}
&\partial_t n + \partial_x J = 0 \, ,\\
&\partial_t J + J = - \partial_x n \, ,
\end{split}
\end{equation}
which is the Cattaneo theory of diffusion. Combining these two equations yields the second-order evolution equation $\partial_t n = (\partial_x^2 - \partial_t^2)n$, as expected.

A natural question is therefore which form of the scattering integral gives rise to the conversion rate $\mathcal{R}=(n_- - n_+)/2$. The simplest example is provided by the relaxation-time approximation, $(\partial_t {+} \V \partial_x)f {=} f_{\mathrm{eq}} {-} f$ \cite{AndersonWitting1974}, where $f_{\mathrm{eq}}$ denotes the local equilibrium distribution with the same density as $f$ (and with the temperature and velocity of the environment). Indeed, integrating this equation over positive momenta yields $\mathcal{R}=n/2 - n_+ = (n_- - n_+)/2$, as required.
By contrast, Fokker-Planck kinetic theory does not share this property. Integrating $(\partial_t + \V \partial_x)f = \beta^{-2}\partial_p[\feq \partial_p (f/\feq)]$ over positive momenta gives
\begin{equation}
\mathcal{R}
=
-\left[\dfrac{\feq}{2\pi\beta^2} \partial_p (f/\feq)\right]_{p=0} \, ,
\end{equation}
which is not determined solely by $n_\pm$, but depends on the detailed momentum-space structure of $f$.


\vspace{-0.2cm}
\section{Determination of the cutoff wavenumber}\label{BBB}
\vspace{-0.1cm}

Plugging \eqref{boostwavemode} into $i\omega=k^2$, we obtain $i\gamma(\Tom-v\Tk)=\gamma^2(\Tk-v\Tom)^2$. We seek solutions with $\Tk=\Lambda\in\mathbb{R}$ and $\Tom=\Omega-i/(\gamma v)$, where the real part $\Omega$ is to be determined, and the imaginary part $-(\gamma v)^{-1}$ saturates \eqref{secondbound}. This leaves us with
\begin{equation}\label{bubbu}
i\gamma(\Omega-v\Lambda)+\dfrac{1}{v}=\gamma^2(\Lambda-v\Omega)^2+2i\gamma (\Lambda-v\Omega)-1 \, .   
\end{equation}
Since $\Lambda$ and $\Omega$ are assumed real, we can just set the real and the imaginary parts of \eqref{bubbu} to zero separately. This immediately gives
\begin{equation}
\Lambda-v\Omega=\pm \dfrac{1}{\gamma}\sqrt{1+\dfrac{1}{v} }\, , \qquad \qquad   \Omega=\dfrac{2+v}{1+2v}\Lambda \, .
\end{equation}
Plugging the second equality into the first, and taking the positive solution, we arrive at \eqref{cutoffone}.

\vspace{-0.2cm}
\section{Fourier decomposition of the rest-frame retarded Green function in boosted frames}\label{ccc}
\vspace{-0.1cm}

We have the following (note the change of integration variable $y=\Tt/v-\Tx$):
\begin{equation}
\begin{split}
G(\Tt,\Tk)={}& \int_{\mathbb{R}} \frac{\Theta(\Tt-v\Tx)}{\sqrt{4\pi\gamma(\Tt-v\Tx)}}
\exp\!\left[
-\frac{\gamma(\Tx-v\Tt)^2}{4(\Tt-v\Tx)} -i\Tk \Tx
\right] d\Tx \\
={}& \int_{\mathbb{R}} \frac{\Theta(y)}{\sqrt{4\pi\gamma vy}}
\exp\!\left[
-\frac{\gamma(y-\frac{\Tt}{\gamma^2 v})^2}{4vy} -i\Tk (\Tt/v-y)
\right] dy \\
={}& \dfrac{e^{-i\frac{\Tk}{v}\Tt+\frac{\Tt}{2\gamma v^2}}}{\sqrt{4\pi\gamma v}}\int_{0}^{\infty} \frac{1}{\sqrt{y}}
\exp\!\left[-
\left(\frac{\gamma}{4v}-i\Tk\right) y- \frac{\Tt\,^2}{4
\gamma^3 v^3 y} 
\right] dy \\
={}& \dfrac{e^{-i\frac{\Tk}{v}\Tt+\frac{\Tt}{2\gamma v^2}}}{\sqrt{4\pi\gamma v}} \sqrt{\dfrac{\pi}{\frac{\gamma}{4v}-i\Tk}} \exp\left[-2\sqrt{\left(\frac{\gamma}{4v}-i\Tk\right)\frac{\Tt\,^2}{4
\gamma^3 v^3}} \right] \\
={}& \dfrac{1}{\sqrt{\gamma (\gamma-4iv\Tk)}} \exp\left[-i\Tt\left(\frac{\Tk}{v}+\frac{i}{2\gamma v^2}-\dfrac{|\Tt|}{\Tt}\dfrac{i}{2\gamma v^2}\sqrt{1-\dfrac{4iv\Tk}{\gamma}}\right) \right]\, , \\
\end{split}
\end{equation}
from which we immediately obtain
\begin{equation}
G(\Tt,\Tk)=\dfrac{1}{\sqrt{\gamma (\gamma-4iv\Tk)}} 
\begin{cases}
e^{-i\Tom_-(\Tk)\Tt} \, , & t>0\, ,\\
e^{-i\Tom_+(\Tk)\Tt} \, , & t<0\, ,\\
\end{cases}
\end{equation}
where $\omega_\pm$ are the two dispersion relations of the boosted diffusion equation. In particular, $\omega_-$ is the stable branch (blue curve in figure \ref{fig:Cutoff}), see equation \eqref{stableDispersion}, while $\omega_+$ is the unstable branch (red curve in figure \ref{fig:CutoffScale}).

\bibliography{Biblio}

\label{lastpage}
\end{document}